\documentclass[]{spie}  
\usepackage[]{graphicx}
\usepackage{amsmath}

\title{Attitude determination for balloon-borne experiments} 

\author{
N.N.~Gandilo\supit{a}$^\dagger$, P.A.R.~Ade\supit{b}, M.~Amiri\supit{c}, F.E.~Angil\`e\supit{d}, S.J.~Benton\supit{e}, J.J.~Bock\supit{f}\supit{,g}, J.R.~Bond\supit{h,i}, S.A.~Bryan\supit{j}, H.C.~Chiang\supit{k}, C.R.~Contaldi\supit{l}, B.P.~Crill\supit{f}\supit{,g}, M.J.~Devlin\supit{d}, B.~Dober\supit{d}, O.P.~Dor\'e\supit{f}\supit{,g}, M.~Farhang\supit{a,h}, J.P.~Filippini\supit{f}, L.M.~Fissel\supit{a,m}, A.A.~Fraisse\supit{n}, Y.~Fukui\supit{o}, N.~Galitzki\supit{d}, A.E.~Gambrel\supit{n}, S.~Golwala\supit{f}, J.E.~Gudmundsson\supit{n}, M.~Halpern\supit{c,i}, M.~Hasselfield\supit{p}, G.C.~Hilton\supit{q}, W.A.~Holmes\supit{g}, V.V.~Hristov\supit{f}, K.D.~Irwin\supit{q,r}, W.C.~Jones\supit{n}, Z.D.~Kermish\supit{n}, J.~Klein\supit{d}, A.L.~Korotkov\supit{s}, C.L.~Kuo\supit{r}, C.J.~MacTavish\supit{t}, P.V.~Mason\supit{f}, T.G.~Matthews\supit{m}, K.G.~Megerian\supit{g}, L.~Moncelsi\supit{f}, T.A.~Morford\supit{f}, T.K.~Mroczkowski\supit{f}, J.M.~Nagy\supit{j}, C.B.~Netterfield\supit{a,e,i}, G.~Novak\supit{m}, D.~Nutter\supit{b}, R.~O'Brient\supit{f,g}, E.~Pascale\supit{b}, F.~Poidevin\supit{u,v}, A.S.~Rahlin\supit{n}, C.D.~Reintsema\supit{q}, J.E.~Ruhl\supit{j}, M.C.~Runyan\supit{g}, G.~Savini\supit{w}, D.~Scott\supit{c}, J.A.~Shariff\supit{a}, J.D.~Soler\supit{a,x}, N.E.~Thomas\supit{y}, A.~Trangsrud\supit{g}, M.D.~Truch\supit{d}, C.E.~Tucker\supit{b}, G.S.~Tucker\supit{s}, R.S.~Tucker\supit{f}, A.D.~Turner\supit{g}, D.~Ward-Thompson\supit{z}, A.C.~Weber\supit{g}, D.V.~Wiebe\supit{c}, and E.Y.~Young\supit{n}
\skiplinehalf
\supit{a}Department of Astronomy \& Astrophysics, University of Toronto, Toronto, ON, Canada;\\
\supit{b}School of Physics and Astronomy, Cardiff University, Cardiff, UK;\\
\supit{c}Department of Physics \& Astronomy, University of British Columbia, Vancouver, BC, Canada;\\
\supit{d}Department of Physics \& Astronomy, University of Pennsylvania, Philadelphia, PA, USA;\\
\supit{e}Department of Physics, University of Toronto, Toronto, ON, Canada;\\
\supit{f}Division of Physics, Mathematics \& Astronomy, California Institute of Technology, Pasadena, CA, USA;\\
\supit{g}Jet Propulsion Laboratory, Pasadena, CA, USA;\\
\supit{h}Canadian Institute for Theoretical Astrophysics, Toronto, ON, Canada;\\
\supit{i}Canadian Institute for Advanced Research, Toronto, ON, Canada;\\
\supit{j}Department of Physics, Case Western Reserve University, Cleveland, OH, USA;\\
\supit{k}Astrophysics \& Cosmology Research Unit, University of KwaZulu-Natal, Durban, South Africa;\\
\supit{l}Theoretical Physics, Blackett Laboratory, Imperial College, London, UK;\\
\supit{m}Center for Interdisciplinary Exploration and Research in Astrophysics, Northwestern University, Evanston, IL, USA;\\
\supit{n}Department of Physics, Princeton University, Princeton, NJ, USA;\\
\supit{o}Institute for Advanced Research, Nagoya University, Furo-cho, Chikusa-ku, Nagoya-shi, Japan;\\
\supit{p}Department of Astrophysical Sciences, Princeton University, Princeton, NJ, USA;\\
\supit{q}National Institute of Standards and Technology, Boulder, CO, USA;\\
\supit{r}Department of Physics, Stanford University, Stanford, CA, USA;\\
\supit{s}Department of Physics, Brown University, Providence, RI, USA;\\
\supit{t}Kavli Institute for Cosmology, University of Cambridge, Cambridge, UK;\\
\supit{u}Instituto de Astrof\'isica de Canarias, La Laguna, Tenerife, Spain;\\
\supit{v}Departamento de Astrof\'isica, Universidad de La Laguna, La Laguna, Tenerife, Spain;\\
\supit{w}Department of Physics \& Astronomy, University College London, London, UK;\\
\supit{x}Institut d'Astrophysique Spatiale, CNRS \& Universit\'e Paris-Sud, Orsay, France;\\
\supit{y}Department of Physics, University of Miami, Coral Gables, FL, USA;\\
\supit{z}Jeremiah Horrocks Institute of Maths, Physics and Astronomy, University of Central Lancashire, Preston, UK;\\
}

\authorinfo{N.N.G..: E-mail: gandilo@astro.utoronto.ca, Telephone: 1 416 946 0946\\
Copyright 2014 Society of Photo-Optical Instrumentation Engineers. One print or electronic copy may be made for personal use only. Systematic reproduction and distribution, duplication of any material in this paper for a fee or for commercial purposes, or modification of the content of the paper are prohibited.}

\begin{document} 
  \maketitle 

\begin{abstract}
An attitude determination system for balloon-borne experiments is presented. The system provides pointing information in azimuth and elevation for instruments flying on stratospheric balloons over Antarctica. In-flight attitude is given by the real-time combination of readings from star cameras, a magnetometer, sun sensors, GPS, gyroscopes, tilt sensors and an elevation encoder. Post-flight attitude reconstruction is determined from star camera solutions, interpolated by the gyroscopes using an extended Kalman Filter. The multi-sensor system was employed by the Balloon-borne Large Aperture Submillimeter Telescope for Polarimetry (BLASTPol), an experiment that measures polarized thermal emission from interstellar dust clouds. A similar system was designed for the upcoming flight of \textsc{Spider}, a Cosmic Microwave Background polarization experiment. The pointing requirements for these experiments are discussed, as well as the challenges in designing attitude reconstruction systems for high altitude balloon flights. In the 2010 and 2012 BLASTPol flights from McMurdo Station, Antarctica, the system demonstrated an accuracy of $<5^{\prime}$ rms in-flight, and $<5^{\prime\prime}$ rms post-flight.
\end{abstract}



\section{INTRODUCTION}
\label{sec:intro}

This paper describes an attitude determination system for balloon-borne telescopes that was designed to meet the pointing requirements of two astrophysical experiments: BLASTPol \cite{Pascale} and \textsc{Spider}\cite{Fraisse}. BLASTPol is a submillimeter polarimeter that maps magnetic fields in star-forming regions of the galaxy, and \textsc{Spider} will map the polarization of the Cosmic Microwave Background over $\approx$10\% of the sky. The performance of this system is evaluated using data from the 2010 and 2012 flights of BLASTPol from McMurdo Station, Antarctica. Attitude sensors that will be used on \textsc{Spider}'s first flight in 2014 were also tested on the BLASTPol flights.

\section{SYSTEM OVERVIEW}
\begin{figure} 
\centering
\includegraphics[width=1.0\textwidth]{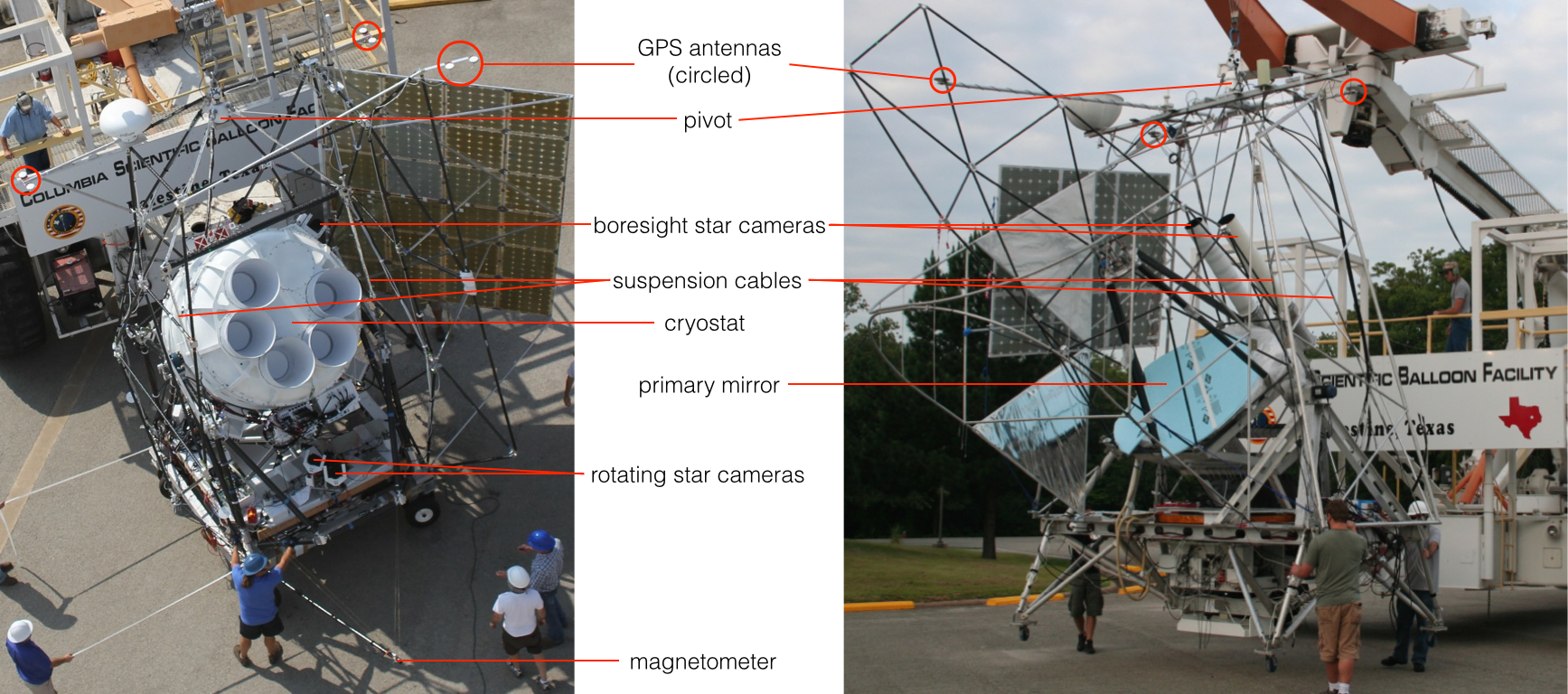}
\caption{Photographs taken during compatibility testing at the Columbia Scientific Balloon Facility. Left: \text{Spider}. Right: BLASTPol.}\label{bothofem}
\end{figure}

Figure \ref{bothofem} shows the overall structure of each experiment, as well as the location of various pointing sensors. Each telescope has a gondola structure\cite{soler2014mecha} that is designed to support the telescope and point it on the sky\cite{jas2014}. Both gondolas have an outer frame that moves in azimuth and an inner frame that moves in elevation. The outer frame is suspended from the pivot by cables. For \textsc{Spider}, the main structural component of the inner frame is the large cryostat which houses the six telescopes. For BLASTPol, the inner frame consists mainly of the primary and secondary mirrors of the telescope, with the cryostat mounted behind the primary mirror.

The attitude determination system is designed to achieve two goals. First, the in-flight pointing needs to be accurate enough to ensure that the detector array observes the desired targets. Second, the post-flight reconstructed pointing solution must be accurate enough to make a map of the region that oversamples the angular resolution of the telescope. The pointing requirements for a given experiment therefore depend on the angular size and resolution of the detectors. For BLASTPol, the detector array size is on the order of $10^\prime$, and the diffraction-limited resolution is $30^{\prime\prime}$. BLASTPol's in-flight pointing is required to have $\sim$30$^{\prime\prime}$ accuracy, and it's post-flight pointing needs $<$5$^{\prime\prime}$ accuracy. \textsc{Spider}'s array size is on the order of $10^\circ$, with a resolution of $0.5^\circ$. \textsc{Spider} requires in-flight pointing of $\sim$1$^\circ$ accuracy, and post-flight pointing of $<$10$^\prime$. 

These goals are achieved by combining information from multiple pointing sensors in real-time. Table \ref{sensors} lists the various sensors used, the rate at which they provide readings, and their expected accuracy. These sensors provide absolute attitude solutions, while integrated velocity data from gyroscopes are used to interpolate between absolute positions. The signals from the sensors are read by the BLASTbus electronics system\cite{benton2014} and used by the flight computer to construct an in-flight pointing solution.

\begin{table}[h]
\caption{Pointing sensors.} 
\label{sensors}
\begin{center}       
\begin{tabular}{|l|c|c|}
\hline
\rule[-1ex]{0pt}{3.5ex}  Sensor & Rate (Hz) & Accuracy ($^\circ$) \\
\hline
\hline
\rule[-1ex]{0pt}{3.5ex}  GPS & 10 & 0.1 \\
\hline
\rule[-1ex]{0pt}{3.5ex}  Sun Sensor & 20 & 0.1 \\
\hline
\rule[-1ex]{0pt}{3.5ex}  Magnetometer & 100 & 5 \\
\hline
\rule[-1ex]{0pt}{3.5ex}  Clinometer & 100 & 0.1 \\
\hline
\rule[-1ex]{0pt}{3.5ex}  Star Camera & 0.5 & $<$0.001 \\
\hline
\rule[-1ex]{0pt}{3.5ex}  Elevation Encoder & 100 & $<$0.01 \\
\hline 
\end{tabular}
\end{center}
\end{table} 

\section{Gyroscopes}
\subsection{Gyro Box Components}
Two identical gyro boxes were built for both BLASTPol and \textsc{Spider}. BLASTPol's gyro box is mounted on the inner frame of the telescope, so it tracks both azimuth and inner frame elevation. \textsc{Spider}'s gyro box is mounted to the floor of the gondola on the outer frame, meaning it primarily measures motion in azimuth. For the purpose of redundancy, two gyroscopes are mounted on each of the three orthogonal axes of motion - pitch (elevation), roll, and yaw (azimuth). The gyros are KVH DSP-3000 digital fiber-optic gyroscopes (FOGs).

The DSP-3000 gyro uses a digital signal processor and an open loop optical circuit of polarization-maintaining fiber. The angle random walk of the gyro is $0.0667^\circ/\sqrt{\mbox{hr}}$ (or $4.0^{\prime\prime}/\sqrt{\mbox{s}}$), which comes from noise in the rate signal.  The angle random walk is proportional to the square root of the integration time and therefore limits the length of time one can rely solely on information from the gyros in the absence of another reference.  This determines the frequency at which solutions from the star camera need to be available. The gyro scale factor error, which is a measure of the linearity in the voltage response to the rotational rate, is 1500ppm (=0.15\%). The gyro also has a bias error, which is the signal output in the absence of any real rotation. This error is $\pm20^{\prime\prime}$/s and slowly varies over time.

The other components of the gyro box are the heaters, readout circuit, power supply and relays. To prevent the gyros from getting cold, five 100$\Omega$ resistors are connected in parallel, and the temperature of the box is measured using a thermistor encased in a thermally conductive epoxy. Each gyro is powered through a relay that allows it to be individually switched. The signals from each gyro are amplified through Schmidtt-trigger logic ICs and sent to the flight computer. 

If there is no update in the reported value from a gyroscope, the flight computer identifies that the gyroscope has failed. It then power cycles the failed gyro while masking its signal. Gyros are also able to have their signals masked or their power cycled by command during the flight.

\subsection{Magnetic Shielding}
Fiber optic gyroscopes are affected by the Earth's magnetic field due to the Faraday effect.  This occurs because the magnetic field causes a rotation of the polarization of the light in the optical fiber, and this results in a phase shift which affects the signal.  In order to protect the gyros from this effect, they are wrapped in Metglas Alloy 2714A, a cobalt-based magnetic shielding foil that is only 0.0006 inches thick and has a magnetic permeability of 1,000,000. The response of the gyros to a magnetic field was seen to decrease by a factor of $\sim$10 when the shielding was applied. 

\subsection{Orthogonalization}
Once the gyroscopes are mounted inside the box they must be orthogonalized.  The orthogonalization process determines the orientation of the gyros with respect to three mutually orthogonal axes. This process was done by placing the gyro box on a rotary table and turning it around three orthogonal axes of rotation.  Let the three unit vectors in this orthogonal reference frame be $\hat{x}$, $\hat{y}$ and $\hat{z}$, and the direction vectors of each gyro be $\hat{1}$, $\hat{2}$, ..., $\hat{6}$.  If the box is rotated at angular velocity $\omega=(\omega_x,\omega_y,\omega_z)$ and the angular velocity measured by each gyro is $g_1, g_2, ... , g_6$, then
\begin{equation}\label{gyros}
\omega_x\hat{x}+\omega_y\hat{y}+\omega_z\hat{z} = g_1\hat{1}+g_2\hat{2}+g_3\hat{3}+g_4\hat{4}+g_5\hat{5}+g_6\hat{6}.
\end{equation}
If the box is rotated in the $x-$axis, then $\omega_y=\omega_z=0$ and the gyro signals are
\begin{equation}
g_i=\omega_x\hat{x}\cdot\hat{\imath}, \mbox{ for } i=1...6.
\end{equation}

Then the ratio of one gyro signal versus another, for example $g_2$ vs. $g_1$, is related to the direction vectors of the gyros by
\begin{equation}
\frac{g_2}{g_1}=\frac{\hat{x}\cdot\hat{2}}{\hat{x}\cdot\hat{1}}.
\end{equation}
Therefore by plotting the gyro signals against each other and performing a linear least-squares fit, and repeating this for the other two axes, the orientation of each gyro with respect to each other is found. The gyro orientation angles are then incorporated into the flight code.

Three sides of the gyro box were designed to be angled at exactly $90^\circ$ to each other, in order to place the box on the rotary stage at three orthogonal orientations.  The slight non-orthogonality of the box sides was measured using a dial indicator, and the results incorporated into the rotation matrix of the gyroscopes. The orthogonalization process was repeated several times, and the orientations of the gyros were found to be consistent to $<$10$^{\prime\prime}$. After a $90^\circ$ scan, this $10^{\prime\prime}$ uncertainty in the orientation of any gyro will result in an uncertainty of $15.7^{\prime\prime}$ in the in-flight pointing solution, which is well within the allowed error.    

Every time the gyro box is mounted on the gondola, its orientation in the reference frame of the gondola must be determined. On BLASTPol, this is done by scanning the gyro box around 3 orthogonal axes of rotation while on the gondola. First the inner frame is pointed at an elevation of zero, and the gondola is scanned in azimuth 20 times at $2^{\circ}$/s. These scans are then repeated with the inner frame at an elevation of $90^{\circ}$. Finally, the inner frame is scanned up and down in elevation 20 times at $2^{\circ}$/s. Each of the three final gyroscope signals, representing the 3 axes of rotation, is then a combination of all 6 gyro signals, rotated into the reference frame of the gyro box. 

\section{Star Cameras}
Star cameras provide much more accurate attitude readings than the coarse sensors, although the coarse sensors are typically simpler to operate. Antarctic balloon flights occur during the summer when there is 24-hour daylight. Therefore, star cameras must be able to detect stars above the daytime background level of atmospheric brightness at stratospheric altitudes, which is on the order of a few nW/sr/cm$^2$/nm for wavelengths $>$600 nm.  

BLASTPol uses a pair of star cameras\cite{rex} mounted above the inner frame (Figure \ref{bothofem}) as its main in-flight pointing sensors. These star cameras have operated on BLAST\cite{pascale07} since its 2003 test flight from Fort Sumner, New Mexico, with slight modifications to the hardware. The cameras take 80ms exposures every $\sim$1s, while the gondola moves in azimuth at $0.1^\circ/$s, and 180ms exposures at scan turnarounds when the gondola is stationary.

\textsc{Spider} has two star cameras mounted on the gondola outer frame, and one boresight camera on the cryostat (Figure \ref{bothofem}).  The outer frame cameras sit on a rotating platform at the front of the gondola floor. These rotating star cameras stay fixed on the sky as the gondola scans sinusoidally back and forth in azimuth, with a peak azimuthal velocity of $6^{\circ}$/s.  The rotating cameras take 300ms exposures every $\sim$4 seconds. The boresight camera takes pictures at every scan turnaround, when the gondola is stationary in azimuth, and before the inner frame is stepped in elevation. One of the \textsc{Spider} star cameras has been test-flown on BLASTPol, both in 2010 and 2012. It was bolted to the outer frame of the gondola, and took exposures at every scan turnaround.

\subsection{CCD Camera}
The cameras used by the two experiments are listed in Table \ref{tab:cams}. 
BLASTPol uses QImaging Retiga-EXL CCD cameras. The advantage of these cameras is their high speed readout, allowing up to 15 frames per second to be read out over FireWire. The camera also has increased sensitivity in the near-infrared, which is the part of the spectrum where atmospheric brightness drops off, allowing higher signal-to-noise detection of stars. Interface with the camera is over Windows using the QImaging QCam driver.

The \textsc{Spider} cameras are ST-1603 ME Santa Barbara Instrument Group (SBIG) CCD cameras. The cameras were chosen for their high well depth, allowing for longer exposures without saturating. Communication with the camera is over USB 2.0, with a full frame download time of 2s. Because the \textsc{Spider} star cameras do not have the sky moving across their field of view, the exposures can be longer. They can also be sampled less frequently than the BLASTPol cameras since the pointing requires less accuracy. The ST-1603 ME also has high quantum efficiency that extends into the near-infrared. The linux based SBIG Universal Driver Library is used to send commands to the camera.

\begin{table}[h]
\caption{Star camera specifications.} 
\label{tab:cams}
\begin{center}
\begin{tabular}{|c|c|c|}
\hline
& BLASTPol & \textsc{Spider} \\
\hline
\hline
Camera & Retiga-EXL & SBIG ST-1603ME\\
CCD & Sony ICX285 &Kodak KAF-1603ME \\
Pixel Size & $6.45 \times 6.45\mu$m & $9 \times 9\mu$m \\
Pixel Array & $1392 \times 1040$ pixels & $1530 \times 1020$ pixels\\
Read Noise & 6.5e- & 18e- \\
Well Depth & 16,000e-& $\sim100,000$e-\\
Digital Output & 14 bits &16 bits\\
\hline
\end{tabular}
\end{center}
\end{table}
\subsection{Optics}
BLASTPol uses a Nikon lens with a 200 mm/f2.0 focal length, giving a $2^\circ \times 2.5^\circ$ field of view and a pixel scale of $7^{\prime\prime}$. The \textsc{Spider} camera is coupled with a Sigma 120-300mm f/2.8 telephoto lens. A 200 mm focal length gives the camera a $4^{\circ} \times 2.7^{\circ}$ field of view and a pixel scale of 9.3$^{\prime\prime}$. In order to reduce the background level due to sky brightness, both experiments use filters to cut off wavelengths shorter than 600nm.

During ascent, large temperature fluctuations cause the focus of the lens to change. Therefore a mechanism is needed to be able to focus the camera once the balloon has reached float altitude. BLASTPol uses a stepper motor attached to a belt to move the focusing ring of the lens. \textsc{Spider} uses a simpler method, attaching an Electronic Lens Interface (ELI) made by Birger Engineering to the Sigma lens. The ELI mimics the interface of a Canon camera, allowing the star camera computer to control the lens over USB.

\subsection{Baffles}
\begin{figure} 
\centering
\includegraphics[width=60mm]{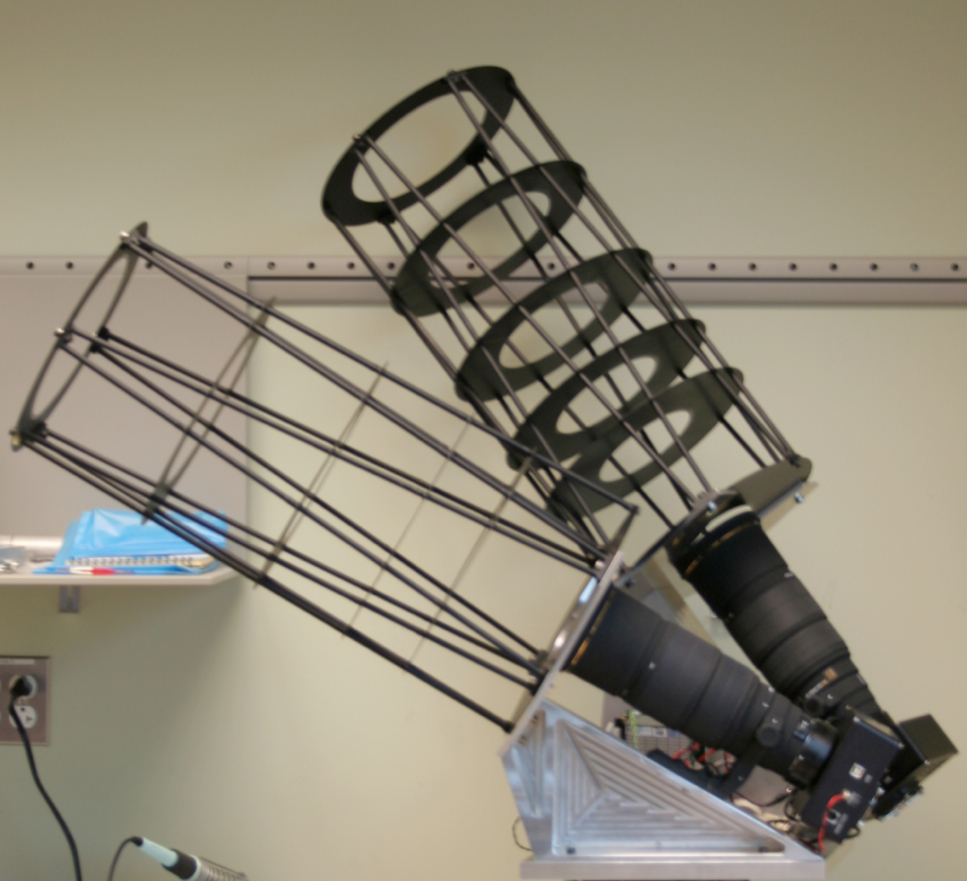}
\caption{Rotating \textsc{Spider} star cameras. The baffles mounted in front of the lenses will later be wrapped in aluminized Mylar.}\label{rmounts}
\end{figure}

In order to block stray light, each star camera lens has a baffle mounted in front of it. Both experiments use knife-edge aluminum disks, with the spacing of the disks designed to eliminate primary reflections from sources $>$10$^{\circ}$ from the optical axis. BLASTPol uses a G-10 tube to hold the disks in place. \textsc{Spider} has a much tighter mass budget, so the disks are held in place by a lightweight carbon fiber truss (Figure \ref{rmounts}), made of Easton Lightspeed size 500 arrow shafts, cut to the desired length. The shafts are slotted through holes in the disks and glued in place with ScotchWeld adhesive epoxy. The carbon fiber baffle truss is then wrapped in aluminized mylar. The insides of the baffles are painted black to minimize internal reflections.

\subsection{Computer and Software}
Each camera has its own PC-104 single board computer, booting from a solid state drive. The computer runs software that controls both the camera and the lens, and communicates with the flight computers over ethernet. The software downloads images from the CCD, and then processes them in order to identify stars. 

Both experiments use similar procedures to locate stars in the star camera image. First the mean and standard deviation of the image are calculated. Sources (``blobs") are identified as pixels that are above a signal-to-noise threshold, and also have neighbouring bright pixels (to rule out single bright pixels). The flux across several pixels is interpolated, in order to determine the centre of the blob.

The star camera computer also runs a separate program that displays the images of the star camera field. The images are relayed to the ground as an NTSC signal by video transmitters during the early part of the flight when line-of-sight communication is possible. The images show useful information such as boxes around sources identified as stars, labeled in order of flux magnitude, as well as a timestamp of when the image was taken. The BLASTPol video display also contains the pointing solution and stellar magnitudes for identified stars. The line-of-sight video is highly useful for initial evaluation of star camera performance at float, and allows for a visual confirmation while initially focusing the cameras. Images throughout the flight are periodically saved to disk, and can be used in post-flight analysis to identify stars that were missed in flight.

Every time an image is taken, the star camera computer sends an ethernet packet to the flight computers. It contains the $x,y$ positions of the 15 brightest blobs, their fluxes, the mean and rms intensity of the image, the focus position of the lens, the CCD temperature, the time the image was taken, and the duration of the exposure. For BLASTPol, an electrical pulse is sent to the star cameras to trigger an exposure. The \textsc{Spider} cameras cannot be externally triggered by an electrical pulse, but the exposure command sent to the boresight camera is timed so that it occurs when the gondola is at the scan turnaround.

The Pyramid algorithm \cite{mortari} is used to identify the stars observed by the star camera blob-finder. It is a `lost-in-space' algorithm, meaning a search is conducted over the entire sky. A faster algorithm was developed for the BLASTPol star cameras, which uses the approximate pointing provided by coarse sensors and only searches a reduced part of the sky around that position. This was needed because BLASTPol's in-flight pointing requires the star cameras to identify stars every 2s, and a lost-in-space search would be too time consuming. Once the stars are identified, the pointing solution reported for that image comprises the celestial coordinates in Right Ascenscion (RA) and Declination (Dec) of the centre of the image, as well as the CCD rotation angle on the sky. The \textsc{Spider} in-flight pointing requirements can be satisfied by coarse sensors, so the star cameras do not need to provide in-flight pointing solutions. However the \textsc{Spider} cameras take pictures infrequently enough that a full Pyramid search can be performed if necessary to allow a pointing solution to be calculated and ensure that the cameras are working properly.

\subsection{Rotating Mount}
\textsc{Spider}'s peak scan velocity is much faster than BLASTPol's ($6^\circ$/s vs. $0.2^\circ$/s), so to prevent stars from streaking across the images, cameras are mounted to a rotating platform on the floor of the gondola outer frame (Figure \ref{bothofem}). Each camera is attached to a lens and mounted onto a triangular aluminum base, as shown in Figure \ref{rmounts}. Mounting the two cameras at different angles, both in azimuth and elevation, allows the pointing solution from one camera to constrain the roll of the other camera. The rotating cameras are mounted to an aluminum platform on an Aerotech ADRT 150-135 direct drive rotating stage. The motor is controlled using a Technosoft IDM-240. The motor speed is commanded over RS-232 serial by the flight computers, and a 16-bit encoder in the motor measures its position to an accuracy of $4^{\prime\prime}$. While the star cameras are exposing, the stage rotates at a speed to counter the azimuthal rotation of the gondola, as reported by the gyros. The full width of \textsc{Spider}'s scans is $90^\circ$, however the opening at the front of the gondola frame through which the cameras can see is much less than $90^\circ$. Therefore after each exposure, the rotary stage returns the cameras to a forward facing position. Mechanical limit switches prevent the cameras from rotating too far in either direction where they might interfere with the cryostat.

\section{Coarse Pointing Sensors}
Both BLASTPol and \textsc{Spider} use a combination of coarse sensors to determine attitude in flight. The purpose of the coarse sensors on BLASTPol is to orient the payload roughly in position. The coarse attitude is a starting point around which the star cameras can search a catalogue to identify stars. For \textsc{Spider}, the in-flight pointing solution relies only on the coarse sensors. Coarse sensors are also used to provide an estimate of the slowly-varying offset in the gyros.

\subsection{GPS}
The GPS measures the azimuth, pitch, and roll of the gondola outer frame. In addition to attitude, the GPS provides time, position (latitude, longitude, altitude), and speed (horizontal and vertical). Two GPS antennas are sufficient to determine the position of the payload, but for attitude determination three antennas are required: a main antenna and two auxiliary antennas. 

BLASTPol's 2010 flight used a Septentrio PolaRx2e@ GPS. The main antenna is a Septentrio PolaNt* dual frequency GPS antenna, operating at $1575\pm10$MHz and  $1227\pm10$MHz. The auxiliary antennas are PolaNt*\_SF single frequency antennas, operating at $1575\pm2$MHz. The receiver communicates with the flight computer over RS-232 serial in a proprietary Septentrio Binary Format (SBF). The settings on the receiver are configured to enable SBF output, set the type of output packets (time, position, and attitude) and set the output rate to 10Hz. The receiver's primary communication port (COM1) outputs the messages, while the secondary port (COM2) can be used on the ground to send commands and debug the unit. These settings are saved as a configuration file in non-volatile memory, forcing the receiver to start up in that configuration after a device reboot. The GPS receiver time is also used as a Network Time Protocol (NTP) server to synchronize the clocks of the two redundant flight computers, through an NTP daemon running on the flight computers. The star camera computers also run their own NTP daemons, and synchronize their clocks to whichever flight computer is currently in charge.

BLASTPol's 2012 flight acquired attitude information from the GPS unit flown by the Columbia Scientific Balloon Facility (CSBF). GPS data is used by CSBF's Support Instrumentation Package (SIP), which provides commanding and telemetry capabilities to the gondola during the flight. The CSBF GPS is a Thales Navigation ADU5, which outputs position and attitude messages in National Marine Electronics Association (NMEA) format at 1Hz. The flight computer is connected over RS-232 serial to a secondary communication port on the ADU5, with the primary communication port being used by CSBF.

The accuracy of the GPS attitude depends on properly choosing the placement of the GPS antennas. First, the distance between the antennas should be as large as possible, given the size of the payload. Second, the angle between the main-auxiliary1 and main-auxiliary2 baselines should be as close to $90^\circ$ as possible. This is to maximize the projection of the main-auxiliary2 baseline onto the plane perpendicular to the main-auxiliary1 baseline, which maximizes the accuracy of the roll determination. Third, the antennas should be placed as far away as possible from any reflective surfaces on the payload (e.g. large antennas, mylarized sunshields). This is to reduce the effect of multipath error, which is the largest source of error in GPS attitude. Multipath refers to the error introduced by the signal from a GPS satellite being reflected by some surface before reaching the antenna.

For the 2010 BLASTPol flight, the Septentrio antennas were bolted to flat plates at the ends of aluminum tubes (see Figure \ref{bothofem}). Two 0.5m tubes were attached to the top back corners of the sunshields, pointing directly outward. A third 3m pole was cantilevered off the top of the sunshields, pointing forward. This produced an L-shaped antenna array, with a 3m main-auxiliary1 baseline, a 3m main-auxiliary baseline, and $90^\circ$ between the main-auxiliary1 and main-auxiliary2 baselines. For this antenna spacing, the expected accuracy is $0.1^\circ$, $0.2^\circ$, and $0.2^\circ$ in heading, pitch, and roll, respectively. For the 2012 flight, the CSBF GPS antennas were mounted on top of the sunshields with a 1m separation between antennas, providing an accuracy of $0.3^\circ$ in heading and $0.6^\circ$ in pitch and roll. \textsc{Spider} will use a Septentrio receiver in its 2014 flight, with two antennas attached to the back corners of the sunshields, pointing outward, and one antenna at the tip of a forward-facing side wing of the sunshields (see Figure \ref{bothofem}).

When the Septentrio GPS was first mounted onto the BLASTPol gondola in 2010, the antennas were not near any reflective surfaces, and the calibration and testing showed no problems. However the forward most antenna was mounted on a long boom that placed it next to the carbon fiber inner frame baffle structure (see Figure \ref{bothofem}). Later in the campaign this structure was covered in aluminized Mylar, creating a highly reflective surface near the front GPS antenna. At that point the mounting of the GPS antennas could not be changed, so the multipath effect is visible in the resulting performance. 

\subsection{Pinhole Sun Sensors}
Both BLASTPol and \textsc{Spider} use multiple pinhole sun sensors \cite{Korotkov} to determine the outer frame azimuth based on the position of the sun. The sun sensors use Hamamatsu S5991-01 two-dimensional 9$\times$9mm position-sensitive detectors (PSDs). A PSD consists of a uniform P-type resistive layer formed on the surface of a N-type high-resistivity silicon substrate. The P-layer acts as an active area for photoelectric conversion and a pair of output electrodes are formed on both ends of the P-layer for extracting signals. A common electrode is connected to the backside of the silicon substrate.

When a light spot hits the PSD, it generates an electric charge that is proportional to the light intensity. The electric charge travels through the resistive layer to the output electrodes, generating a photocurrent that is inversely proportional to the distance between the position of the incident light and the electrode. The output is then read through an operational amplifier, with the output voltage determined by the feedback resistance and the input current. The required output voltage is between 5-10V in order to be within the dynamic range of the BLASTbus analog readout.

\begin{figure} 
\centering
\includegraphics[width=0.36\textwidth]{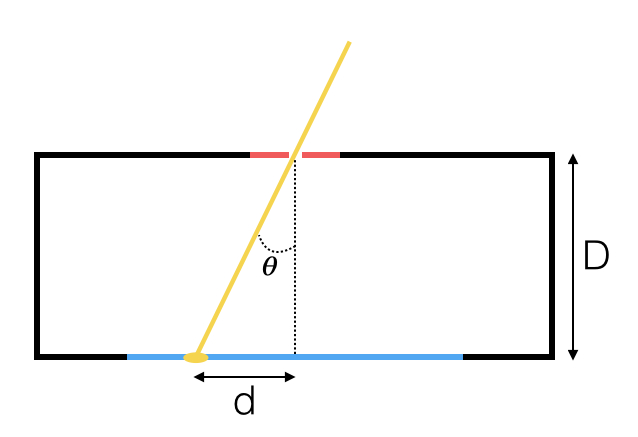}
\caption{Light passing through the pinhole and hitting the PSD. The angle, $\theta$, between the sensor and the sun can be calculated from the distance, d, between the light spot and the centre of the PSD, and the distance, D, between the sensor and the pinhole.}\label{mypssfig}
\end{figure}

Sunlight enters through a $200\mu$m copper precision pinhole from Edmund Optics, mounted 10mm above the centre of the active area of the PSD. Once the position of the light spot is known, the angle between the sensor and the sun can be calculated, knowing the distance from the sensor to the pinhole (see Figure \ref{mypssfig}). Only detections of a light spot $\leq$4mm from the centre of the sensor are considered, since the accuracy degrades near the edge of the sensor. This gives each sun sensor a field of view of $2\times\arctan\frac{4\mathrm{mm}}{10\mathrm{mm}}\simeq40^\circ$.  

\begin{figure} 
\centering
\includegraphics[width=75mm]{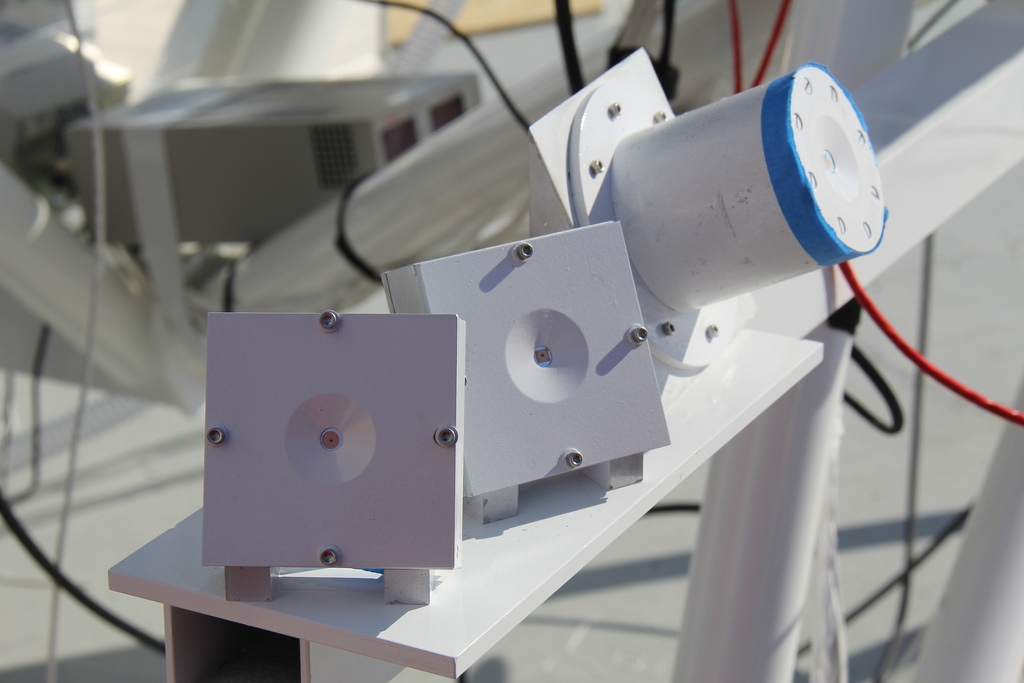}
\caption{Pinhole sun sensors.}\label{pssboom}
\end{figure}

Each sensor is held in an aluminum case, with the inside painted black. The sensors are mounted at different angles in azimuth in order to cover the range over which the gondola scans in azimuth relative to the sun. The angles are also chosen so that the sensors overlap each other's field of view by $\sim$2$^\circ$. The elevation of the sun varies from $15-35^\circ$ in Antarctica in December, so each case is mounted at $25^\circ$ elevation. The sensors are mounted on a boom attached to bottom of the outer frame of the gondola, in a spot that ensures no other parts of the gondola cast shade on the sensors (Figure \ref{pssboom}). The outer frame is preferred over the top of the sunshields as a mounting point because it is more rigid and less susceptible to warping, which makes the orientation of the sensors more fixed over the course of the flight.

\subsection{Magnetometer}
A magnetometer provides coarse azimuth of the outer frame by detecting changes in the direction of the Earth's magnetic field. The direction of the magnetic field lines at any given location is given by the National Oceanic and Atmospheric Administration (NOAA) World Magnetic Model. The model only accounts for the portion of the magnetic field generated in the Earth's fluid outer core. It does not include effects generated in the crust and upper mantle, or in the magnetosphere and ionosphere. The model is only updated every 5 years, so the model has the highest accuracy when it is released, and subsequently degrades over the next 5 years. The current model was published in December 2009, and will expire in December 2014.

Near the south pole, the magnetic field lines are highly inclined, so detecting changes in the component parallel to the horizon is less effective. For Antarctic balloon experiments like BLASTPol and \textsc{Spider}, the magnetometer is the least useful sensor and contributes negligibly to the in-flight pointing solution whenever more accurate sensors are functioning. For balloon flights at lower latitudes, the accuracy of the magnetometer is expected to be much better.

The 2012 BLASTPol flight used a Honeywell HMR2300 three-axis digital magnetometer, which is also the model being flown on \textsc{Spider}. It uses three magneto-resistive sensors (HMC 1001) to measure the orthogonal X,Y and Z components of the magnetic field. The magneto-resistive sensors use a thin film of Permalloy (nickel-iron) deposited on silicon as a resistive strip. The Permalloy is magnetized in a particular direction, and its resistance to current depends on the angle between the direction of magnetization and the direction of current flow. In the presence of the Earth's magnetic field, the magnetization direction is deflected, causing a change in resistance. This produces a corresponding change in voltage across the output of a wheatstone bridge. The HMR2300 has an internal analog-to-digital converter, and outputs in serial RS-485. A RS-485 to RS-232 serial converter was used to communicate with the flight computers. The magnetometer is configured to output messages in ASCII format, with a continuous output rate of 20Hz. Configuration settings are stored by an onboard EEPROM. The magnetometer is mounted on a long boom attached to the most forward point of the gondola (Figure \ref{bothofem}), in order to prevent magnetic interference from the payload.

\subsection{Clinometer}
BLASTPol uses two biaxial clinometers from Applied Geomechanics. Both are variants of the 904-T ``Clinometer Pak". One clinometer is mounted to the bottom of the gondola, to measure pitch and roll of the outer frame. It is the Model 904-TH high-gain version, with a $\pm10^\circ$ span. The second clinometer is mounted to the inner frame as an inner frame elevation sensor. It is the Model 904-TS standard version, and has a $\pm25^\circ$ span. The clinometer uses a liquid-filled electrolytic transducer as the sensing element. The sensor cannot detect pendulations of the gondola, only constant tilt about two orthogonal tilt axes. Its response to tilt also has a temperature dependency. Overall it provides an accuracy of $\sim$0.1$^\circ$.

\textsc{Spider} has two clinometers, both Model 904-TH. One is mounted in the control box of the pivot motor, which is the connection point between the payload and the flight train of the balloon. The control box is attached to the bottom face of the pivot, so the clinometer measures the pitch and roll of the pivot. The second clinometer is mounted on the floor of the gondola, and measures the pitch and roll of the outer frame of the gondola. 

\subsection{Elevation encoder}
Elevation of the inner frame relative to the outer frame is provided by absolute encoders. BLASTPol uses a Kollmorgen C053A as an elevation motor, which has a built-in encoder with sub-arcsecond accuracy, although it is read out with $20^{\prime\prime}$ resolution.  \textsc{Spider} has two EA 58-S 16-bit electro-optical encoders mounted on either side of the cryostat, measuring elevation to an accuracy of $0.0055^\circ$. Because the encoders measure the elevation angle of the inner frame relative to the outer frame, they do not account for pendulations in the outer frame. Therefore they do not measure the absolute elevation of the telescope boresight, and this limits their accuracy as elevation sensors to $\sim$0.1$^\circ$.

\section{In-flight pointing solution}
A separate pointing solution is calculated for each sensor. It is a weighted sum of the sensor's old pointing solution, $p_{old}$, and the new reading from the sensor, $x$:
\begin{equation}
p=\frac{w_1p_{old} + w_2x}{w_1+w_2} + g + t
\end{equation}
where $w_1=\sigma^2_{sys}$ is the systematic variance of the sensor and $w_2=w_\mathrm{samp}$ is the sample weight given to each new reading of the sensor. These two weights are defined according to the accuracy of the sensor. $g$ is the change in angle given by integrating the gyros over the time since the sensor's last reading, and $t$ is a trim angle that can be added by command to adjust the sensor's solution. The variance in the pointing solution is then given by
\begin{equation}
\sigma^2=w_1+\frac{1}{w_1+w_2}.
\end{equation}

The separate solutions for each sensor, $p_i$, then get combined into a pointing solution, $P$:
\begin{equation}
P=\frac{\sum(\sigma^2_i+w_{1,i})p_i}{\sum(\sigma^2_i+w_{1,i})},
\end{equation}
A separate $P_{\mathrm{AZ}}$ and $P_{\mathrm{EL}}$ are calculated by combining either azimuth or elevation sensor readings. A sensor can also be vetoed by command during the flight if it appears to be malfunctioning. All un-vetoed sensors are used to calculate an estimate of the offset in the gyroscopes, which slowly varies over time. 

\section{POINTING RECONSTRUCTION}
The post-flight reconstructed pointing solution is based solely on data from the star cameras and the gyroscopes, using a method developed for BLAST.
\subsection{Gyroscope rotation}
In order to use the gyroscope data to integrate between star camera solutions, the rotation from the gyro reference frame to the star camera reference frame must be calculated. This can be done using data from the flight where large slews were done with good star camera solutions before and after the slew. Starting from the star camera solution at the beginning of each slew, a pointing solution at the end of the slew is calculated based solely on integration of gyroscope rate data. This gyroscope-based pointing solution differs from the star camera solution at the end of the slew. A least-squares fit is then done to calculate the rotation that minimizes this difference over all slews.

\subsection{Star camera solutions}
The star camera data is recorded by the flight computers at a later time than when the camera exposure was taken. Therefore, before the star camera data can be used to reconstruct a pointing solution for the flight, it needs to be synchronized in time.  The trigger pulse can be used to pinpoint the exposure time, and produce a set of time-synchronized star camera fields. In cases where the trigger pulse did not register, the star camera data can be synchronized based on the exposure time reported by the star camera in microseconds. 

Pointing solutions are then calculated based on reported blob positions for the entire flight. In flight, a minimum of 3 stars must be matched before a pointing solution is accepted. Post-flight, this minimum is reduced to 1. At first, the in-flight pointing solution is used as the starting point to match stars, but once a gyro-integrated pointing solution is calculated, this can be used instead. As the accuracy of the gyro-integrated pointing solution increases, more 1- and 2-star solutions can be found.

\begin{figure}
\centering
\includegraphics[width=.55\textwidth]{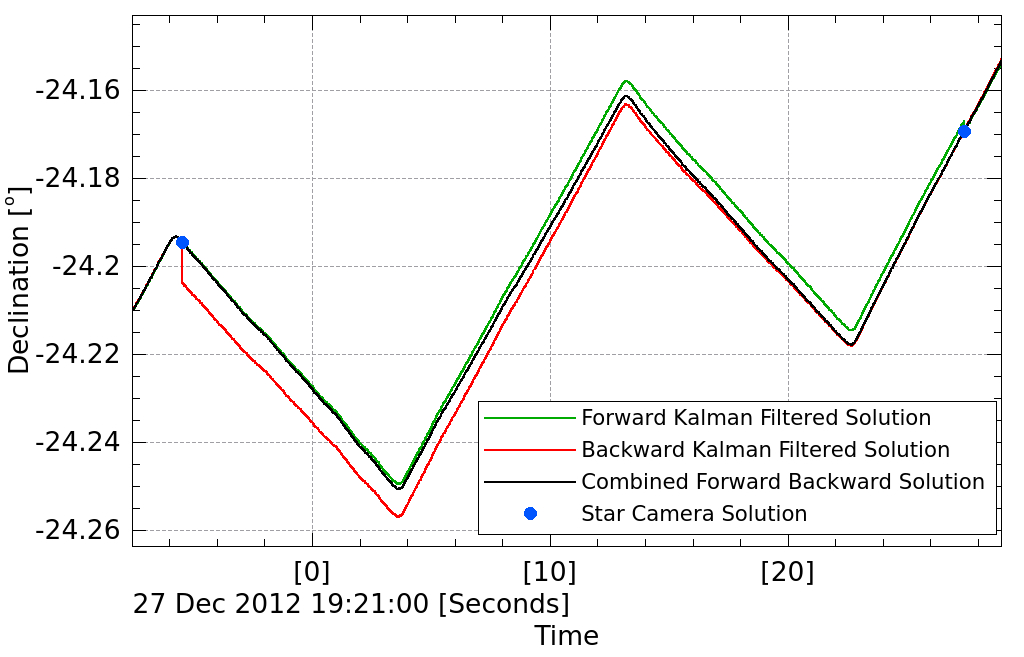}
\caption{Kalman Filtered pointing solution.}\label{kalman}
\end{figure}

\subsection{Kalman integration}
The rotated gyroscope timestreams are used to interpolate between each star camera solution. The pointing solution algorithm employs an Extended Kalman Filter method used by the Wilkinson Microwave Anisotropy Probe (WMAP)\cite{harman05}. The attitude of the telescope at each point in time is represented as a quaternion, which can be written as a set of 4 numbers $q=(w,x,y,z)$ where $w$ is real and $x,y,z$ are imaginary. In this representation, $w=cos(\theta/2)$, where $\theta$ is the amount of rotation about the axis defined by $(x,y,z)$. A quaternion can also be written as a scalar and a vector $q=(s,\textbf{v})$, with $s=w$ and $\textbf{v}=(x,y,z)$. At each time step $\Delta T=10$ms, a quaternion multiplication is performed to rotate the attitude of the telescope by the rotation quaternion given by the gyroscopes
\begin{equation}
q_{n+1}=
\begin{pmatrix} 
1  \\ 
\frac{1}{2}w_n\Delta T
\end{pmatrix}
q_n +
\begin{pmatrix} 
0  \\ 
\frac{1}{2}\textbf{b}_n\Delta T
\end{pmatrix}
q_n +
\begin{pmatrix} 
0  \\ 
\frac{1}{2}\textbf{p}_{w_n}\Delta T
\end{pmatrix}
q_n
\end{equation}

where $\textbf{b}_n$ is the gyro bias and $\textbf{p}_{w_n}$ is the noise in the gyro rate measurement.

At each point in time where there is a star camera solution, the error in the solution jumps to zero and then subsequently grows with the integrated gyro error until the next star camera solution. The filter is run both forward and backward in time, and the forward and backward solutions are then combined, weighted by their respective errors. This smooths the solution over the periods in which there are no star camera solutions (Figure \ref{kalman}).

\subsection{Results}
\begin{figure}
\centering
\includegraphics[width=.49\textwidth]{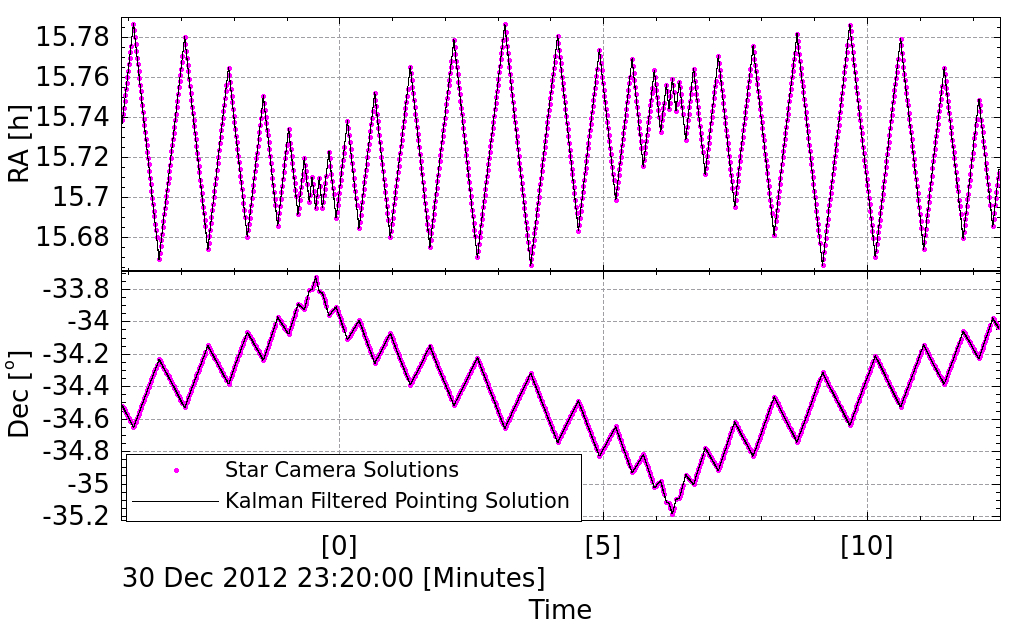}
\includegraphics[width=.49\textwidth]{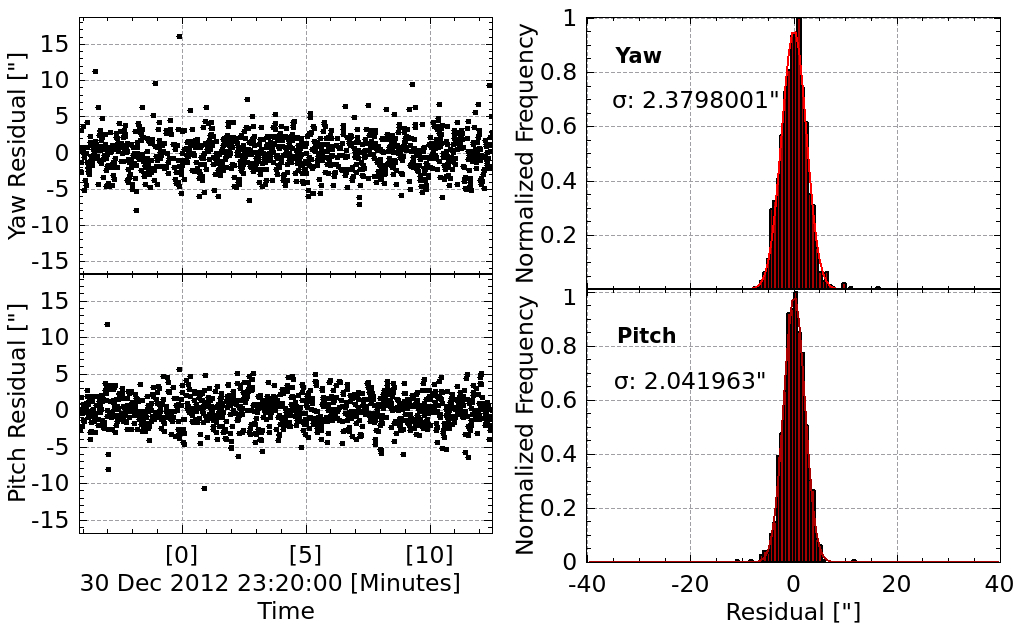}
\caption{Left: Points are star camera solutions and solid line is the final Kalman filtered pointing solution. Right: Residuals in the pointing solution in pitch and yaw.}\label{residual}
\end{figure}

Figure \ref{residual} shows the star camera solutions and the Kalman filtered pointing solution in RA and Dec for $\sim$15 minutes of the BLASTPol 2012 flight. The residuals in yaw (=RA$\times$cos(Dec)) and pitch (=Dec) are also shown. These are the difference between the Kalman filtered solution and the star camera solutions. The sum in quadrature of the rms residual in pitch and yaw is the error in the pointing solution, which is $\sim$3$^{\prime\prime}$. 

\section{Flight Performance}
\subsection{Coarse sensors}
\begin{figure}
\centering
\includegraphics[width=.45\textwidth]{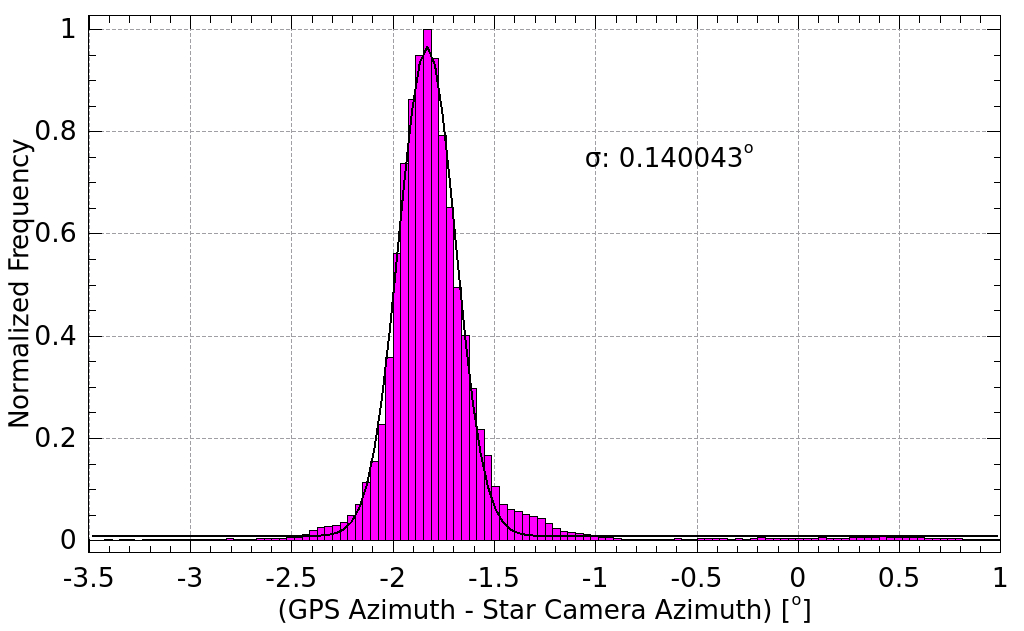}
\includegraphics[width=.45\textwidth]{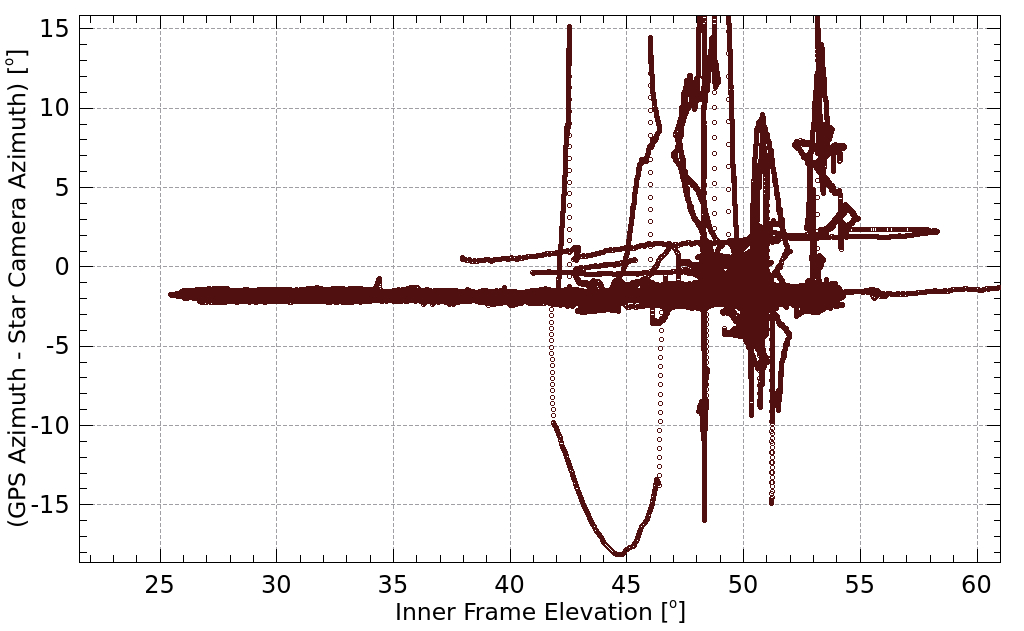}
\caption{Left: Accuracy of the azimuth provided by the Septentrio GPS in the 2010 BLASTPol flight. Right: Dependence of the azimuth accuracy on the elevation of the inner frame, showing large errors above $40^\circ$ elevation.}\label{gps}
\end{figure}

\begin{figure}
\centering
\includegraphics[width=.45\textwidth]{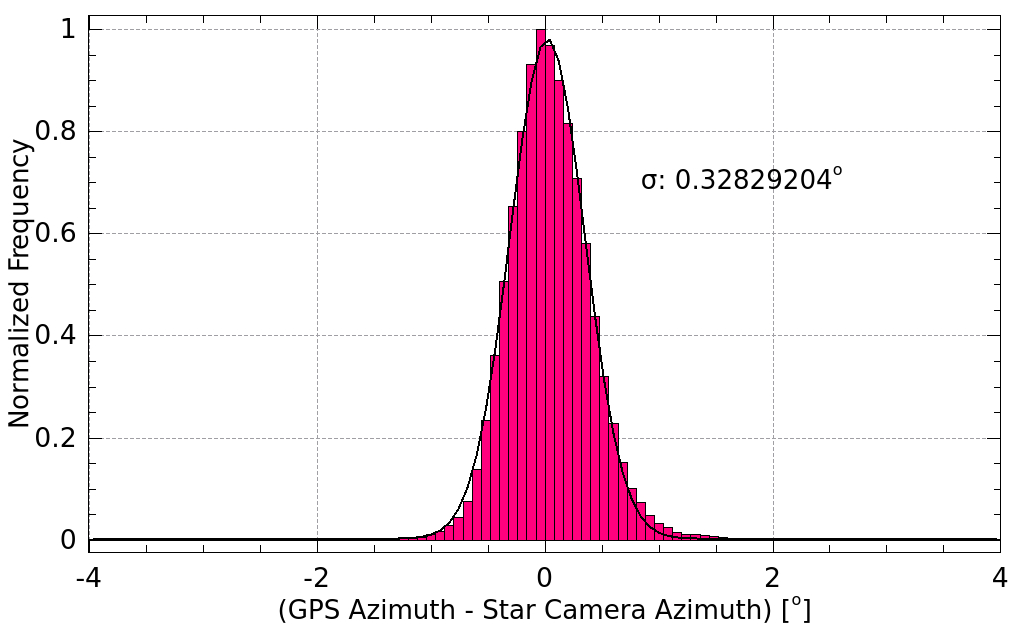}
\includegraphics[width=.45\textwidth]{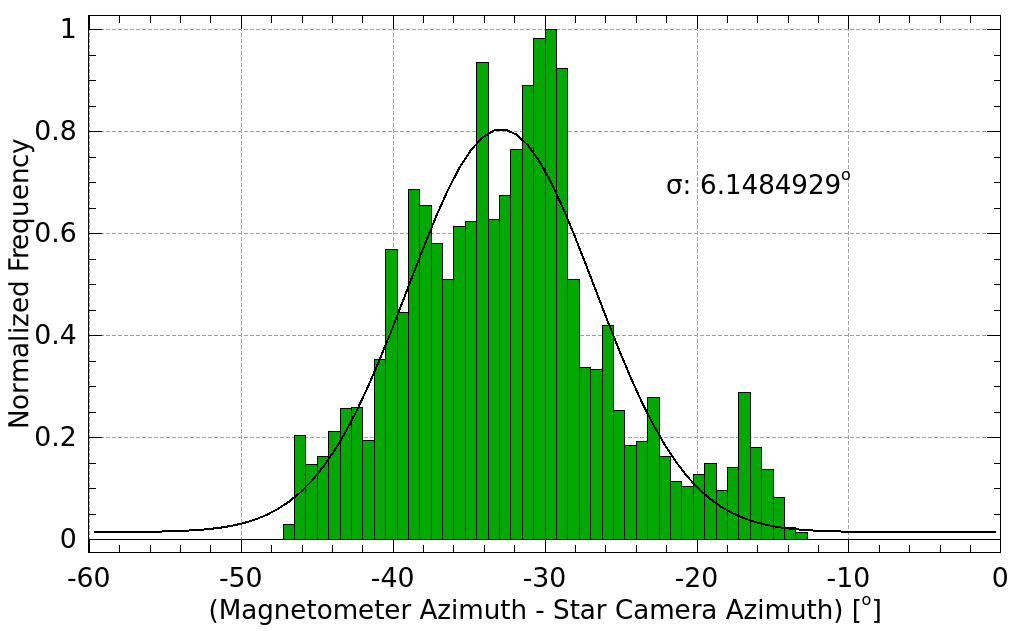}
\includegraphics[width=.45\textwidth]{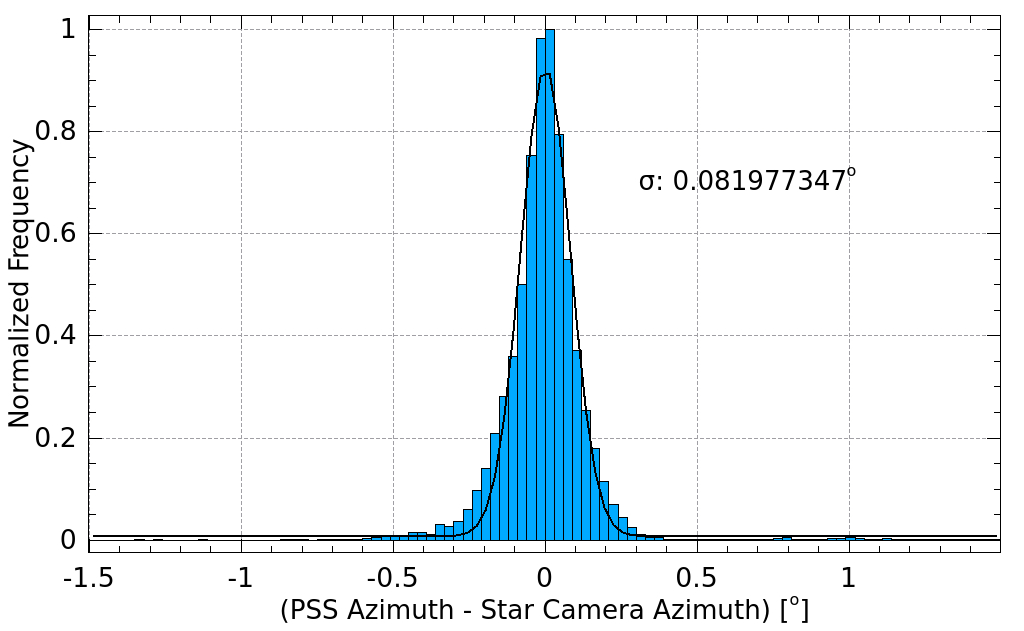}
\includegraphics[width=.45\textwidth]{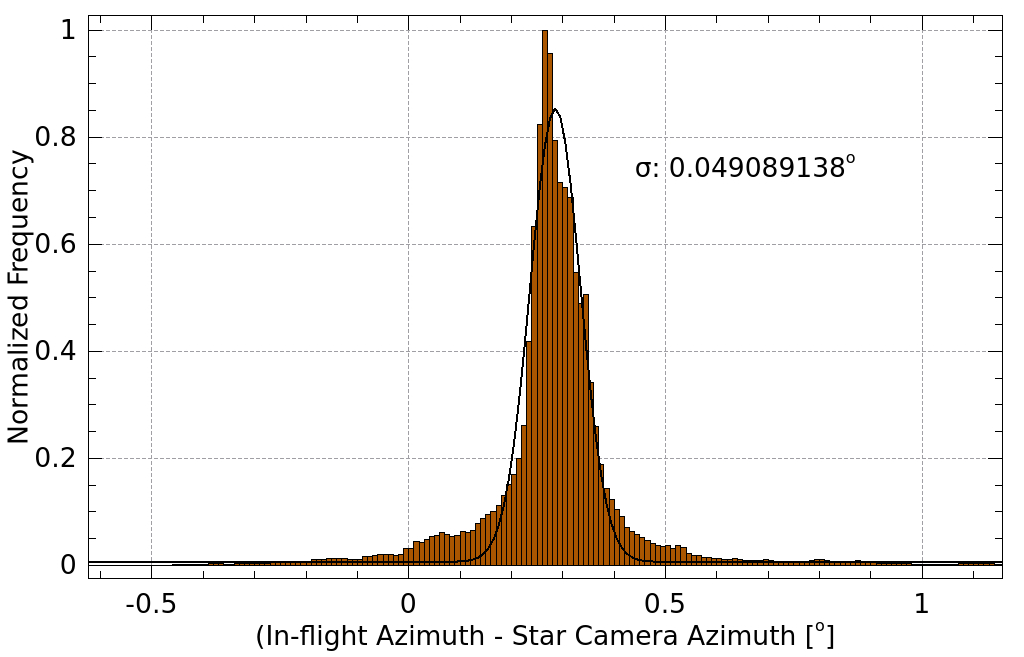}
\caption{Accuracy of coarse sensors and combined in-flight pointing solution. Top Left: ADU5 GPS. Top Right: Magnetometer. Bottom Left: Pinhole Sun Sensor. Bottom Right: In-flight combined azimuth.}\label{coarses}
\end{figure}

After the flight, the performance of each coarse pointing sensor is evaluated by analyzing the difference between the raw azimuth provided by the sensor and the reconstructed star camera-based azimuth. Figures \ref{gps} and \ref{coarses} show histograms of the differences between the coarse sensors and the star camera. For each sensor, a Gaussian is fit to the histogram, and the full-width at half maximum of the fit is taken as the accuracy of the sensor.

Both GPS units performed as expected given the size of their antenna arrays. The performance of the Septentrio GPS on the 2010 BLASTPol flight was sometimes affected by the inner frame baffle of the telescope. After being shielded in Mylar, the baffle became a large reflective surface, resulting in large GPS error any time the inner frame was pointed above $40^\circ$ in elevation. The multipath effect is clearly visible when looking at the dependence of the GPS accuracy on the elevation of the inner frame (Figure \ref{gps}). When not being blocked by the inner frame baffle, the attitude was accurate to $\sim$0.14$^\circ$ rms. The azimuth measured by the ADU5 in the BLASTPol 2012 flight differed from the star camera azimuth by $\sim$0.33$^\circ$ rms (Figure \ref{coarses}).  

The performance of the other coarse sensors is shown in Figure \ref{coarses}, as well as the combined in-flight pointing solution azimuth. As expected, the magnetometer produced the least accurate azimuth reading, with an rms accuracy of $\sim$6$^\circ$. The accuracy of the pinhole sun sensors can be improved post-flight by adjusting various parameters that affect the azimuth calculation for each sensor. These parameters are defined in-flight according to design specifications and pre-flight calibration tests, but their actual value can be more accurately determined post-flight by fitting to the star camera azimuth. The parameters include the elevation mounting angle of the sensors, the roll angle of the sensor, the distance between the pinhole and the sensor, and the azimuth mounting angle offset between the telescope boresight and the sensor. After post-flight calibration, the accuracy of the pinhole sun sensors was found to be $\sim$0.08$^\circ$ rms. The accuracy of the combined in-flight pointing solution is $\sim$0.05$^\circ$ rms.

\subsection{Star cameras}
\subsubsection{Hard drive failures}
On the BLASTPol 2012 flight, both boresight star cameras suffered failures of their hard drives, the first one failing within 6 hours of launch, and the second one after 6 days. The cause of the failure is unknown, but the only drives that failed on the flight were those used by the star camera, and both were the same model (Intel 320). All other drives on the flight were of a different model, and none failed, including the \textsc{Spider} star camera drive and the flight computer drives. As a result of the failure of the boresight star cameras, the second half of the flight ($\sim$6 days) will use a pointing solution based on the \textsc{Spider} star camera mounted on the outer frame.
\subsubsection{Mesospheric clouds}
\begin{figure}
\centering
\includegraphics[width=.66\textwidth]{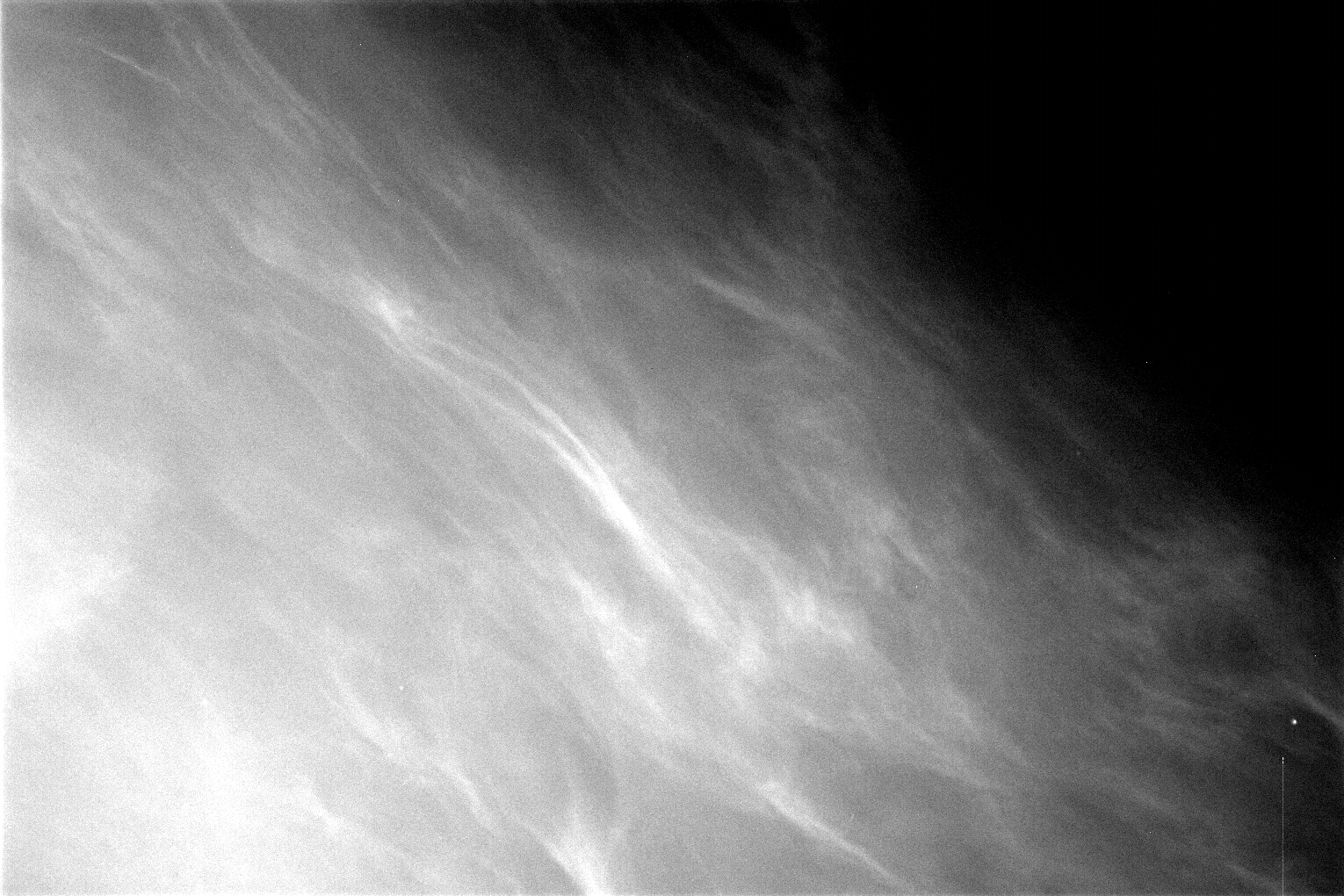}
\caption{Image from the star camera of cloud-like structures observed during the BLASTPol 2012 flight.}\label{clouds}
\end{figure}

All three star cameras flown in the BLASTPol 2012 flight observed structures on the sky (Figure \ref{clouds}) which are believed to be clouds formed from ice crystals in the mesosphere. The presence of the structures interfered with the star camera's ability to detect stars, but they are only visible in a small fraction of the images that were saved during the flight.
\subsubsection{BLASTPol star cameras}
Before suffering hard drive failures, the BLASTPol star cameras performed well. As described in Section 5, the resulting pointing solution obtained using data from the BLASTPol star camera over the first 6 days of the flight has an accuracy of a few arcseconds.
\subsubsection{\textsc{Spider} star camera}
\begin{figure}
\centering
\includegraphics[width=.45\textwidth]{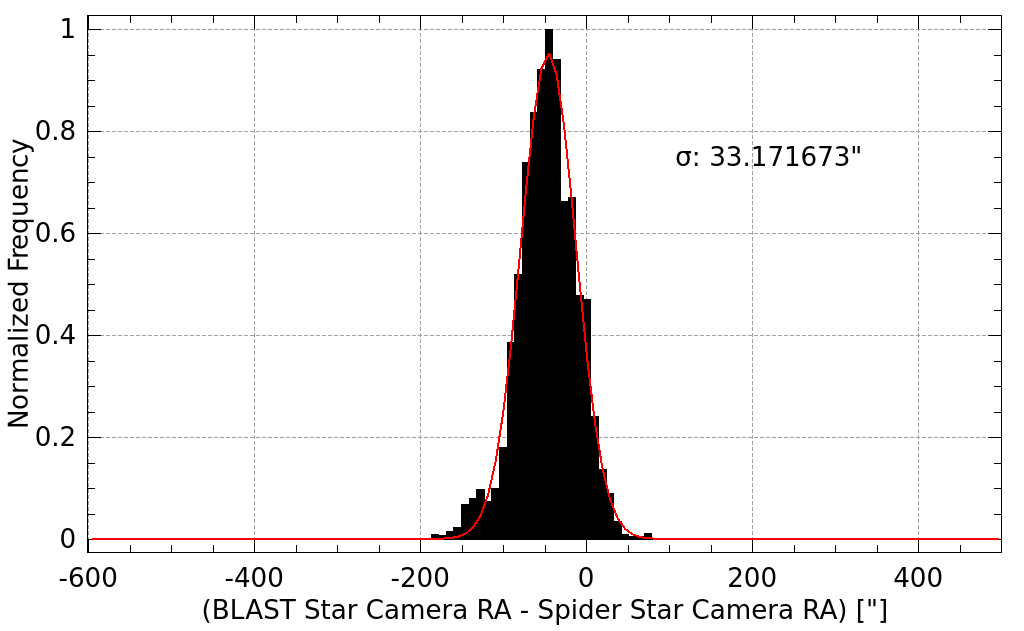}
\includegraphics[width=.45\textwidth]{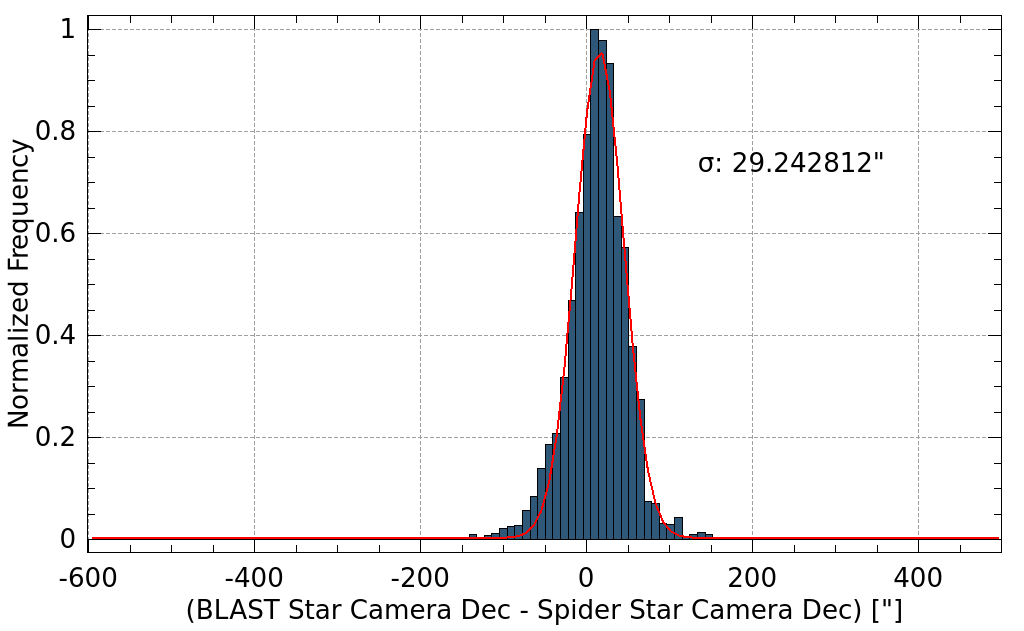}
\caption{Residual uncertainty in the rotation of the \textsc{Spider} star camera solution into the boresight reference frame.}\label{rolly}
\end{figure}

The \textsc{Spider} star camera test flown on the BLASTPol 2012 flight was able to operate successfully and detect stars for the full duration of the flight. A pointing solution was reconstructed from the \textsc{Spider} star camera data using the procedure described in Section 5. Since the camera was mounted on the outer frame of the BLASTPol gondola, its solution had to be rotated into the reference frame of the telescope boresight, which is on the inner frame. This required using coarse sensors to determine the attitude of the outer frame relative to the inner frame. The residual uncertainty in this rotation was $\sim$30$^{\prime\prime}$ rms, which was determined by comparing the rotated \textsc{Spider} star camera solution to the BLASTPol boresight star camera solution (Figure \ref{rolly}) during a period of the flight when both were active.

\section{CONCLUSIONS}
An attitude reconstruction system has been designed which provides the necessary in-flight and post-flight pointing accuracy for balloon-borne astrophysical observations. The system has been developed over the course of several Antarctic flights, including the 2006 flight of BLAST and the 2010 and 2012 flights of BLASTPol. The system is capable of providing in-flight pointing accuracy of $<$5$^\prime$ and post-flight pointing accuracy of $<$5$^{\prime\prime}$ using star cameras and gyroscopes. Coarse sensors (pinhole sun sensors and GPS) are able to achieve accuracy of $0.1^\circ$ in azimuth. Although both BLASTPol star cameras suffered hard drive failures in the 2012 flight, a \textsc{Spider} camera being test flown was able to produce a post-flight pointing solution that could be rotated into the boresight reference frame to $\sim$30$^{\prime\prime}$ accuracy. 

\acknowledgments     
 
The BLAST collaboration acknowledges the support of NASA through grant numbers NNX13AE50G S03 and NNX09AB98G and the Leverhulme Trust through the Research Project Grant F/00 407/BN. The \textsc{Spider} collaboration acknowledges the support of NASA through grant numbers NNX07AL64G and NNX12AE95G. We acknowledge the support of the Lucille and David Packard Foundation, the Gordon and Betty Moore Foundation, the Natural Sciences and Engineering Research Council (NSERC), the Canadian Space Agency (CSA), the Canada Foundation for Innovation, the Ontario Innovation Trust, the Fondo Institucional para la Investigacion of the University of Puerto Rico, the Rhode Island Space Grant Consortium, and the National Science Foundation Office of Polar Programs. We thank the JPL Research and Technology Development Fund for advancing detector focal plane technology.  W.~C.~Jones acknowledges the support of the Alfred P. Sloan Foundation. A.~S.~Rahlin is partially supported through NASAs NESSF Program (12-ASTRO12R-004). J.~D. Soler acknowledges the support of the European Research Council under the European Union's Seventh Framework Programme FP7/2007-2013/ERC grant agreement number 267934. F. Poidevin thanks the Spanish Ministry of Economy and Competitiveness (MINECO) under the Consolider-Ingenio project CSD2010-00064 (EPI: Exploring the Physics of Inflation) for its support.

Logistical support for this project in Antarctica is provided by the U.S. National Science Foundation through the U.S. Antarctic Program. We would also like to thank the Columbia Scientific Balloon Facility (CSBF) staff for their continued outstanding work.


\bibliography{report}   
\bibliographystyle{spiebib}   

\end{document}